# Spin-Scattering Rates in Metallic Thin Films Measured by Ferromagnetic Resonance Damping Enhanced by Spin-Pumping

C. T. Boone, J. M. Shaw, H. T. Nembach, T. J. Silva

National Institute of Standards and Technology, Boulder, CO 80305

**Abstract**

We determined the spin-transport properties of Pd and Pt thin films by measuring the increase in ferromagnetic resonance damping due to spin-pumping in ferromagnetic (FM)-nonferromagnetic metal (NM) multilayers with varying NM thicknesses. The increase in damping with NM thickness depends strongly on both the spin- and charge-transport properties of the NM, as modeled by diffusion equations that include both momentum- and spin-scattering parameters. We use the analytical solution to the spin-diffusion equations to obtain spin-diffusion lengths for Pt and Pd. By measuring the dependence of conductivity on NM thickness, we correlate the charge- and spin-transport parameters, and validate the applicability of various models for momentum-scattering and spin-scattering rates in these systems: constant, inverse-proportional (Dyakanov-Perel), and linear-proportional (Elliot-Yafet). We confirm previous reports that the spin-scattering time appears to be shorter than the momentum scattering time in Pt, and the Dyakanov-Perel-like model is the best fit to the data.

## I. INTRODUCTION

Spin-transport is currently a topic of enormous interest. New devices based on spin-transport are frequently proposed; such devices utilize phenomena such as pure spin-currents [1], the spin Hall effect [2,3,4, 5, 6, 7], as well as giant- and tunneling-magnetoresistance [8, 9, 10]. A fundamental understanding of how spins propagate in metals, as well as how they are absorbed and transmitted at NM/FM interfaces, is vital for the exploitation of phenomena such as spin-pumping [11, 12,13, 14, 15, 16] and spin-torque from the spin Hall effect (SHE) [17 ,18, 19, 20, 21, 22, 23,24] for technological applications such as magnetic random-access memory, magnetic data storage, and spin-based logic. The effects of charge scattering, proximity-induced polarization at the NM/FM interface, and magnetic "dead" layers on spin-diffusion are subjects of ongoing debate.

In this work, we use a spectroscopic method to address how the momentum- and spin-scattering rates for electrons in metals with strong spin-orbit coupling, e.g., Pt and Pd, affect the diffusive transport of pure spin-currents. We find strong evidence to support the conjecture that momentum- and spin-scattering are indeed coupled processes in thin-films of Pt and Pd, though the coupling process is more subtle than would be presumed from the simple Elliot-Yafet picture for spin-flip processes that affect charge carriers.

An outstanding question remains as to how spin- and charge-transport are related in high-$Z$ metals such as Pt and Pd [25,26,27,28,29,30]. In the standard spin-charge diffusion equations, spin-transport is dependent on the charge conductivity, as can be seen in the standard time-independent, steady-state equations that describe spin-diffusion [31, 32, 33, 34,35],

$$\vec{Q}_{\hat{s}} = -\frac{\hbar\sigma}{2e^2}\vec{\nabla}\mu_{\hat{s}},$$

$$\nabla^2\mu_{\hat{s}} = \frac{1}{\lambda_{sf}^2}\mu_{\hat{s}},$$

(1)

where $\vec{Q}_{\hat{s}}$ is the spin-current dyadic, $\mu_{\hat{s}}$ is spin-accumulation for spins polarized along the $\hat{s}$ direction, $\sigma$ is the electrical conductivity, $\hbar$ is the reduced Planck's constant, and $\lambda_{sf}$ is the spin diffusion length. Here, we have omitted the precessional term of the form $\hat{s} \times \vec{H}$, where $\vec{H}$ is the net magnetic field, because spin-depolarization in metals occurs on much faster timescales (fs) than spin-precession (ps).

The spin-diffusion length $\lambda_{sf}$ and electron mean free path $\ell$ are often treated as independent for the purposes of fitting experimental data [19, 30, 36, 37], but theory suggests they are related. In particular, the Eliot-Yafet (E-Y) scattering mechanism [38, 39], whereby each momentum-scattering event has a certain probability $P$ of being a spin-scattering event, suggests that the spin-flip time is given by $\tau_{sf} = \tau/P$, where $\tau$ is the momentum-scattering time.

The complementary relation between spin- and charge-scattering is the Dyakanov-Perel (D-P) mechanism [40, 41], whereby spins continuously de-phase due to the combination of spin-orbit coupling and crystal-lattice inversion-symmetry breaking until a momentum-scattering event occurs. In this picture, the spin-flip time is inversely proportional to the momentum-scattering time, i.e., $\tau_{sf} \propto \tau^{-1}$. Strictly speaking, the D-P mechanism is not operative in materials with cubic symmetry because they are inversion-symmetric. However, the recent work of Jiao and Bauer [42] found that the conventional theories for spin-pumping, spin-diffusion, and the spin-Hall effect could be successfully employed to fit dc voltages due to spin-pumping and the inverse spin-Hall effect in Permalloy/Pt bilayers for varying Pt thickness if they used measured values for thickness-dependent conductivity $\sigma(L)$, but kept a constant value of $\lambda_{sf} = 1.3$ nm. A constant spin diffusion length with thickness-varying conductivity also explained recent spin Hall magnetoresistance data [43]. The invocation of a constant $\lambda_{sf}$ but varying $\sigma$ has a subtle but important logical consequence; if we invoke the diffusion relation $\lambda_{sf} = \sqrt{D\tau_{sf}}$, where $D$ is the diffusion constant, the Einstein relation $D = (\sigma/e^2 v)$, where $v$ is the density of states, and the Drude model for conductivity $\sigma = (ne^2\tau/m^*)$, where $n$ is the conduction electron density and $m^*$ is the effective mass, we then obtain

$$\tau_{sf} = \left(\frac{\lambda_{sf}^2 m^* v}{n}\right)\frac{1}{\tau}. \qquad (2)$$

Hence, the fitting procedure used in Ref. [42] is phenomenologically equivalent to D-P insofar as it implies that $\tau_{sf} \propto 1/\tau$. At face value, the implication of the procedure used in Ref. [42] would appear to explain various experimental data that suggest $\lambda_{sf} < \ell$ for samples where the NM thickness is comparable to, or even smaller than, the bulk-value for $\ell$ [36, 44]. If $\sigma$ and, therefore, $\tau$ are sufficiently reduced at small NM thicknesses, then it is possible that $\tau < \tau_{sf}$, if $\tau < \sqrt{(\lambda_{sf}^2 m^* v/n)}$. This would satisfy an important requirement for the application of diffusion theory for spin transport in nonferromagnetic metals. In principle, both D-P and E-Y can exist for different scattering sites, leading to a more complicated relation between $\tau$ and $\tau_{sf}$ [45].

The reported data for $\lambda_{sf}$ in Pt and Pd span a significant range (0.5 to 14 nm for Pt [19, 46, 47], 2 to 12 nm for Pd [36, 44, 48, 49, 50], with several reported values smaller than the commonly-assumed mean-free-path for these materials. Because most measurements are performed on sputtered thin films, it has been suggested that the strong thickness- and growth-dependence of the conductivity and interfacial properties for thin-films could explain such a wide range of reported values [51]. Much of the previous work had been performed at low temperature, making extrapolation to room temperature difficult. An alternative explanation rests on the fact that Pt and Pd are "almost" ferromagnetic according to Stoner theory [52]. As such, Pt and Pd are magnetically polarized when in direct contact with a ferromagnet [53, 54], which could affect the spin-diffusion length, and even the functional form of spin-transport [44, 50].

We use the phenomenon of spin-pumping to determine diffusive spin-current flow in metallic multilayers: when a FM is adjacent to a NM metal and the magnetization is out of thermal equilibrium, spin-accumulation forms at the FM/NM interface as a result of spin-pumping [32, 34]. As Fig. 1(a) illustrates, the spin-accumulation diffuses away from the interface, creating a spin-current $\vec{Q}_{\hat{s}}$, which decays with a characteristic length $\lambda_{sf}$, as described by Eq. (1). This spatial flow of angular momentum manifests itself as an increase in the Gilbert damping parameter of the FM [11, 32, 55]. We measure the change in Gilbert damping as a function of layer thicknesses in ferromagnetic multilayers in order to infer the spin-current flowing away from the FM/NM interface. Fitting of the spin-pumping data vs. film thickness allows us to extract $\lambda_{sf}$. This room-temperature, spectroscopic method does not require patterning or electrically conductive contacts and therefore is free of artifacts related to edge defects and contact resistances.

By determining both $\lambda_{sf}$ and $\sigma$ for the same samples, we can ascertain whether there is an interdependence of $\tau_{sf}$ and $\tau$, thereby testing various models for the spin-scattering. In so doing, we present experimental evidence that $\lambda_{sf} < \ell$ for all but the smallest NM thicknesses, and that the thickness-dependence of $\sigma$ cannot resolve this apparent conundrum.

We study spin-transport with two different general categories of multilayers: (1) those with either the NM = Pt or Pd in direct contact with the FM = $Ni_{80}Fe_{20}$ (Permalloy, "Py"); and (2) those with Cu inserted between the NM and the FM. We find that the spin absorption properties for a NM in direct contact with a FM and with a spacer are different; it appears that the polarization of the NM at the interface affects the spin-transport. This suggests that the usual spin/charge diffusion equations may require further augmentation to adequately describe transport in systems where the exchange splitting of the bands is not purely local.

## II. THEORY

When a FM, with instantaneous magnetization unit vector $\hat{m}$, is adjacent to a NM, magnetization dynamics in the FM effectively act as a transverse spin-potential source, so that $2\mu_{\hat{s}}(0) = \hbar \hat{m} \times (d\hat{m}/dt) - \frac{2e^2}{\hbar \text{Re}[G^{\uparrow\downarrow}]} \vec{Q}_{\hat{s}}(0)$ at the FM/NM interface [32]. Spin current of polarization $\hat{s}$ is thus sourced at the FM/NM interface through a conductance $2\text{Re}[G^{\uparrow\downarrow}]$, beyond which it obeys Eq. (1). This is shown diagrammatically in Fig. 1a.

For uniform excitation of the magnetization, the problem reduces to the one-dimensional case with an *N*-layer NM stack adjacent to the FM layer. We can treat each layer *i*, with conductivity $\sigma_i$, thickness $L_i$, and spin diffusion length $\lambda_{sf,i}$, as a spin-resistor ladder network in series with a spin-resistor at the FM/NM interface (i.e., the reciprocal of the spin-mixing conductance $G^{\uparrow\downarrow}$), shown in Fig. 1b, where the two parallel-resistors-to-ground $R_i^p$ for the *i*th layer have the value

$R_i^p = (\lambda_{sf,i}/\sigma_i)\coth(L_i/2\lambda_{sf,i})$ and the series-resistor $R_i^s$ has the value

$R_i^s = (\lambda_{sf,i}/\sigma_i)\sinh(L_i/\lambda_{sf,i})$. Thus, elementary circuit theory allows us to relate the spin-current $Q_{\hat{s}}^0$ at the FM/NM interface and the transverse spin-potential in the FM via matrix multiplication,

$$\begin{pmatrix} 0 \\ \mu_{\hat{s}}^{N} \end{pmatrix} = \Pi \begin{pmatrix} Q_{\hat{s}}^{0} \\ \mu_{\hat{s}}^{0} \end{pmatrix}, \tag{3}$$

where $\mu_{\hat{s}}^{N}$ is the spin-potential at the far edge of the *N*th (i.e., last) layer of the NM stack and

$$\Pi = \begin{bmatrix} \Pi_{1,1} & \Pi_{1,2} \\ \Pi_{2,1} & \Pi_{2,2} \end{bmatrix} = \prod_{i=1}^{N} T_i \tag{4}$$

and

$$T_i = \begin{pmatrix} \cosh\left(\frac{L_i}{\lambda_{sf,i}}\right) & -\frac{\sigma_i}{\lambda_{sf,i}} \sinh\left(\frac{L_i}{\lambda_{sf,i}}\right) \\ -\frac{\lambda_{sf,i}}{\sigma_i} \sinh\left(\frac{L_i}{\lambda_{sf,i}}\right) & \cosh\left(\frac{L_i}{\lambda_{sf,i}}\right) \end{pmatrix}. \tag{5}$$

By use of Eq. (3), we can then solve for the sourced spin-current as

$$Q_{\hat{s}}^{0} = \frac{2G^{\uparrow\downarrow}}{1 + \left(\frac{2G^{\uparrow\downarrow}}{G_{ext}}\right)} \frac{\hbar^2}{4e^2} \hat{m} \times (d\hat{m}/dt), \tag{6}$$

where

$$G_{ext} = -\frac{\Pi_{1,2}}{\Pi_{1,1}}. \qquad (7)$$

The ratio $2G^{\uparrow\downarrow}/G_{ext}$ is sometimes referred to as the backflow factor [34, 36]. The outlined matrix formalism can also accommodate interfacial spin-flip between any two layers by insertion of a fictitious *j*th layer between any two layers to represent any interface, with the substitutions $\delta_j = L_j/\lambda_{sf,j}$ and $AR_j = L_j/\sigma_j = \delta_j\left[\lambda_{sf,j}/\sigma_j\right]$, where $\delta_j$ is the interfacial spin-flip parameter, and $AR_j$ is the interfacial resistivity, as defined in Ref. [56].

The flow of transverse spin-current across the FM/NM interface is an additional source of damping $\Delta\alpha$ for the magnetization dynamics in the FM, which is computed via the general relation

$$\Delta\alpha = \frac{|\gamma|\hbar^2}{4e^2 M_s L_{FM}} \mathrm{Re}\left[2G_{eff}\right];$$
$$2G_{eff} = \frac{2G^{\uparrow\downarrow}}{1+\left(2G^{\uparrow\downarrow}/G_{ext}\right)}, \qquad (8)$$

where $\gamma$ is the gyromagnetic ratio, *e* is the electron charge, $M_s$ is the saturation magnetization of the FM, and $L_{FM}$ is the FM layer thickness. $G_{eff}$ is the effective spin-conductance that restricts the flow of transverse spin-current due to the series arrangement of $G^{\uparrow\downarrow}$ and $G_{ext}$. The combination of Eqs. (3-8) allows us to fit data for the spin-pumping contribution to the damping for any arbitrary *N*-layer NM stack, inclusive of interfacial spin-flipping.

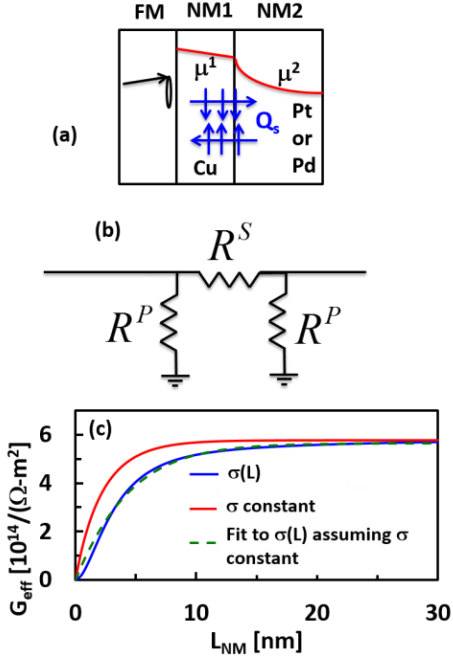

Figure 1: (a) Sketch of the spin-pumping mechanism. (b) Schematic of equivalent ladder-circuit element to describe spin transport for the *i*th layer, from which any multilayer stack can be built. $R^s$ and $R^p$ are equivalent spin resistances that are defined in the main text. (c) Calculated dependence of effective spin-mixing conductance of a FM/NM bilayer for two cases: 1. An unphysical constant conductivity independent of NM thickness (solid red). 2. A realistic NM-thickness-dependent conductivity (solid blue). If we fit the curve for case 2, but erroneously assumes a constant conductivity, the resultant fit is shown as the dashed green curve. While the fit captures the salient qualitative features of the thickness-dependence, the fitted parameters suffer from errors of approximately 50%.

We note that the application of different hypotheses of how NM transport properties are correlated with other film properties can significantly affect the data fits to the diffusive spin-current model, and the wide variability of fitted parameters, such as spin-diffusion length, reported in the literature for materials such as Pt, may be partially explained by this. For example, if the conductivity of a NM layer is assumed to be constant, but is in reality decreasing with decreasing layer thickness, the fitted spin-diffusion length will incur a systematic inaccuracy. Figure 1(c) illustrates how this can happen by comparing calculations, for the illustrative case of a simple FM/NM bilayer, of the expected dependence of $G_{eff}$ on NM thickness, where we consider one material with constant conductivity and another with the experimentally determined thickness-dependent conductivity we observe for Pt, as discussed in more depth later in this paper. Here, we assume $\lambda_{sf}$ = 5 nm and $G^{\uparrow\downarrow} = 10^{15}$ $\Omega^{-1}$ m$^{-2}$. An important point is that a model that assumes a constant value of conductivity provides reasonable fits to the data regardless of whether the conductivity is constant or contains a thickness-dependence. In the case where the simulated data have a thickness-dependent conductivity, the fitted parameters $\lambda_{sf}$ = 7.5 nm and $G^{\uparrow\downarrow} = 1.5 \times 10^{15}$ $\Omega^{-1}$ m$^{-2}$ are 50 % larger than the parameters used to generate the simulated data. Thus, fits of spin-pumping data with a model that ignores the thickness-dependence of the conductivity provide, at best, an *upper bound* on $\lambda_{sf}$ and $G^{\uparrow\downarrow}$.

## III. EXPERIMENT

### A. dc conductivity measurements

To obtain the net dc resistivity $\rho_{\text{NM}}$ of the upper NM layer, we measure the four-probe, current-in-plane sheet-resistance of the multilayer as a function of NM thickness. Interfacial charge scattering results in a resistivity component that scales inversely with NM thickness $L_{\text{NM}}$ [57, 58, 59],

$$\rho_{\text{NM}} = \rho_{\text{b}} + \frac{\rho_{\text{s}}}{L_{\text{NM}}}, \tag{9}$$

where $\rho_{\text{b}}$ is the bulk resistivity, and $\rho_{\text{s}}$ is the interfacial resistivity coefficient. Eq. (9) is numerically consistent with models for the thickness-dependent resistivity of granular thin-films given by solution of the Boltzmann equation, under the assumption of diffusive surface- and grain-boundary scattering [60]. Small deviations from this functional form may occur at the smallest thicknesses. The contribution of all the other layers in the multilayer stack (i.e., the "under-layer") to the resistivity are accounted for by assuming that the under-layer resistivity is constant for a given structure, and that the under-layers together with the NM layer act as parallel resistors so that

$$R_{\text{sheet}} = \left( \frac{L_{\text{UL}}}{\rho_{\text{UL}}} + \frac{L_{\text{NM}}}{\rho_{\text{NM}}} \right)^{-1} \tag{10}$$

where $R_\text{sheet}$ is the sheet resistance of the entire stack, $\rho_\text{UL}$ and $L_\text{UL}$ are the underlayer resistivity and thickness, respectively.

For the purposes of determining the thickness-dependence of NM resistivity in representative multilayer stacks, we sputter-deposited samples of Ta(3 nm)/Py(3 nm)/NM($L_\text{NM}$) onto oxidized Si substrates in a chamber of base pressure of $10^{-7}$ Pa ($10^{-9}$ Torr). Details of the sputtering chamber and representative film roughnesses with Ta seed layers (<0.5 nm) are described in Shaw, et al. [61]. In Figure 2(a), we present the resistivity of a Ta(3 nm)/Py(3 nm)/Pt($L_\text{NM}$) multilayers as a function of $L_\text{NM}$, along with a fit to Eqs. (9) - (10). The model fits the experimental data very well over the entire range of NM-thicknesses that was prepared: 0.5 to 30 nm. The extracted bulk- and surface-resistivities of the NM layers are shown in Table 1. The bulk resistivities that we obtain for Pd and Pt are somewhat higher than tabulated room-temperature values ($1.05 \times 10^{-7}$ Ω m and $1.04 \times 10^{-7}$ Ω m for Pd and Pt, respectively), but are similar to previously-reported measured values of the thin-film resistivity in other studies of the spin diffusion length for Pt ($1.6 \times 10^{-7}$ Ω m for Ref [19]; $1.8 \times 10^{-7}$ Ω m for Ref [37]; $2 \times 10^{-7}$ Ω m for Ref [62]).

| NM | Underlayer | $\rho_b$ ($10^{-7}$ Ω m) | $\rho_s$ ($10^{-16}$ Ω m²) |
|---|---|---|---|
| Pd | Ta(3 nm)/Py(3 nm) | 1.36 ± 0.03 | 4.4 ± 0.3 |
| Pd | Ta(3 nm)/Py(3 nm)/Cu(5 nm) | 1.38 ± 0.04 | 2.1 ± 0.1 |
| Pt | Ta(3 nm)/Py(3 nm) | 1.70 ± 0.03 | 3.3 ± 0.3 |
| Pt | Ta(3 nm)/Py(3 nm)/Cu(5 nm) | 1.71 ± 0.03 | 2.9 ± 0.2 |

Table 1: Resistivities of the Pd and Pt for the structures studied here.

### B. Ferromagnetic resonance

We used ferromagnetic resonance (FMR) to study changes in damping due to spin-pumping from Py into the NMs Pd or Pt. The metals Pd and Pt are of interest because they are "almost" ferromagnetic, i.e., the product of the density of states and the exchange integral almost satisfy the Stoner criterion for itinerant ferromagnetism in Stoner theory [52]. In addition, Pd and Pt are well known to exhibit interfacial spin-polarization when in contact with metallic ferromagnets [53, 54]. Finally, the substantial spin-orbit coupling in these materials gives rise to significant spin-Hall angles [63], making them technologically useful for future spintronic applications. For the FMR measurements, multilayers of Ta(3nm)/Py(3nm)/NM(X) and Ta(3nm)/Py(3nm)/Cu(5nm)/NM(X) were sputter-deposited onto oxidized Si substrates in a chamber with a base pressure of $10^{-7}$ Pa ($10^{-9}$ Torr), and subsequently coated with PMMA. We used a Cu spacer layer to determine the role of interface polarization in spin-pumping measurements. A 5 nm-thick Cu spacer is sufficiently thick to prevent direct exchange-coupling between Permalloy and the NM, but sufficiently thin to prevent loss of spin-accumulation within the Cu [64]. Deposition rates were calibrated with x-ray reflectometry, and thicknesses are accurate to within 2 %, which is included in our estimated error bars. The Ta seed-layer is used to promote a (111) texture of the Permalloy.

Using a standard broadband (5-30 GHz) ferromagnetic resonance technique, we measured for each sample the damping $\alpha$, the spectroscopic $g$-factor $g_L$, and the effective magnetization $M_{\text{eff}}(t_{\text{Py}}) = M_s - (2K_s/\mu_0 M_s t_{\text{Py}})$, where $K_s$ is the net perpendicular interfacial anisotropy with units of J m$^{-2}$, $t_{\text{Py}}$ is the thickness of the Py layer, and $M_s$ is the saturation magnetization [65]. We used an FMR-geometry with the saturating magnetic field applied perpendicular to the sample plane to eliminate two-magnon scattering as an additional source of linewidth broadening [66]. The broadband measurement allows for the accurate extraction of the Gilbert-like component of damping. This geometry also results in precessional motion that is both low amplitude and circularly polarized. Figure 2(b) shows a representative resonance curve, with a fit to the Polder susceptibility that allows for extraction of the resonance field and field-swept linewidth. The extrapolation method developed in Ref. [67] is used to determine $g_L$ and $M_{\text{eff}}$, which are then used for the extraction of $\alpha$ from a fit of the linewidth to $\Delta H = \Delta H_0 + (4\pi \alpha f / [|\gamma| \mu_0])$, where $f$ is the measurement frequency and $\Delta H_0$ is the inhomogeneous linewidth broadening. Representative data for $\Delta H$ vs. $f$ are shown in Figure 2(c). All linewidth data exhibit a purely linear dependence on $f$. By performing FMR measurements with Ta(3nm)/Py($t_{\text{Py}}$)/Pd(10nm) for different Permalloy thicknesses $t_{\text{Py}}$, and subsequent fits to $\alpha = \alpha_0 + (\xi/t_{\text{Py}})$ and $M_{\text{eff}}(t_{\text{Py}})$, we obtain the intrinsic damping $\alpha_0 = 0.0044 \pm 0.0004$ and saturation magnetization $\mu_0 M_s$ = 1.068 ± 0.007 T, in agreement with previously reported values for bulk Permalloy [68]. These results confirm an interfacial component of damping and anisotropy, as previously observed in similar systems [49, 50]. Fig. 2(d) shows the data for the spectroscopic g-factor for all samples with $t_{Py}$ = 3 nm. We find that $g_L$ shows no discernable dependence on NM thickness, and an inconclusive dependence on details of the NM stack, with an average value of $g_L$ = 2.08 ± 0.01.

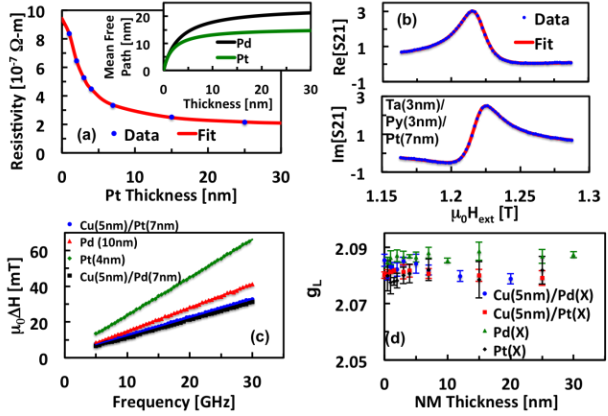

Figure 2: (a) Measured multilayer resistivity as a function of Pt-layer-thickness, with fit to Eqs. (9) and (10) showing the reciprocal-thickness-dependence of the resistivity. Inset shows the extracted mean free paths as a function of thickness. (b) Representative resonance curve, Ta(3nm)/Py(3nm)/Pt(7nm) sample, and fits to the Polder susceptibility, from which the linewidth and resonance field are extracted. (c) Representative linewidths versus frequency, showing the linear dependence across samples, from which α can be precisely extracted (d) g-factor versus NM thickness for all series, showing little variation and no discernable thickness dependence.

## C. Data fitting and analysis

We consider three different fitting-models to interpret our data:

(1) *Both $\tau_{sf}$ and $\tau$ are independent of NM film thickness*. This is equivalent to a constant $\lambda_{sf}$ and a constant bulk-value for σ, similar to prior work. The fitting parameters are $G^{\uparrow\downarrow}$ for the FM/NM interface, $\lambda_{sf}$ for the top NM layer, and $\delta$ for the Cu/NM or the Py/NM interface, depending on whether the sample is a bilayer or trilayer stack.

(2) $\tau_{sf} \propto 1/\tau$. This is equivalent to a constant $\lambda_{sf}$, but a thickness-dependent σ. As discussed in detail earlier, this is identical to the D-P-like model used by Jiao and Bauer to fit inverse spin Hall effect data in Ref. [42]. We use the fitted function of the thickness-dependent $\sigma$ obtained from our four-probe measurements with the aforementioned multilayers. The fitting parameters are the same as in Model 1.

(3) $\tau_{sf} \propto \tau$. This is the E-Y mechanism and is equivalent to $\lambda_{sf} \propto \sigma$, as described earlier. As for Model 2, we use the fitted function of the thickness-dependent $\sigma$. The fitting parameters are $G^{\uparrow\downarrow}$ for the FM/NM interface, $\Lambda = \lambda_{sf}/\ell$ for the top NM layer, and $\delta$ for the Cu/NM or the Py/NM interface, depending on whether the sample is a bilayer or trilayer stack.

For all models, when considering interfacial spin flip, we use $AR = 0.1 \times 10^{-15}$ Ω m² based upon previously reported values for high-quality metallic interfaces [69, 70]. Deviations of AR from this value will affect the exact fitted value of $G^{\uparrow\downarrow}$ but do not affect our other conclusions. Interfacial spin-flip is neglected for the the Py/Cu interface due to the absence of damping enhancement for the case of a FM/Cu/Ta trilayer stack [36], indicative that both the FM/Cu and the Cu/Ta interfaces exhibit negligible spin-flip.

For our 5 nm thick Cu spacer, we measured $\sigma_{Cu}/2L_{Cu}$ =3.6 × 10$^{15}$ ($\Omega^{-1}$ m$^{-2}$). Eqs. (3)-(8) remain valid for $L_{Cu} \to 0$, i.e., no spacer layer, with $G^{\uparrow\downarrow}$, $\delta$, and $AR_I$ changing for the different FM/NM interfaces.

To calculate the momentum-scattering-times $\tau$ and mean-free-paths $\ell$ for Pd and Pt based upon our measured resistivity values, we use the following values obtained from de Haas-van Alphen measurements: a) for Pd, electron density $n$ = 0.376 electrons/atom, average effective mass $m^* = 2.1 m_e$, and average Fermi velocity $v_F = 1.1 \times 10^6$ m s$^{-1}$ [71]; b) for Pt, $n$ = 0.426 electrons/atom, $m^* = 2.4 m_e$, and $v_F = 8.8 \times 10^5$ m s$^{-1}$ [72] and atomic densities $n_{at,Pd}$ = 6.8 × 10$^{22}$ cm$^{-3}$ and $n_{at,Pt}$ = 6.62 × 10$^{22}$ cm$^{-3}$ [73]. Using the Drude equation for resistivity $\tau = m^*/(ne^2\rho)$, we obtain bulk values of $\tau_{Pd}$ = 21 fs and $\tau_{Pt}$ = 18 fs. From the definition of the mean-free-path $\ell = v_F \tau$, we obtain bulk mean-free-paths $\ell_{Pd}$ = 23 nm and $\ell_{Pt}$ = 16 nm.

The results of all fits are tabulated in Table 2. Figures 3(a)-3(d) show the data of $\alpha$ vs. NM thickness $x$ for Py/Cu/Pd($x$), Py/Cu/Pt($x$), Py/Pd($x$), and Py/Pt($x$), respectively, along with fits to Models 1, 2 and 3.

| Multilayer | Model | $G^{\uparrow\downarrow}$ ($10^{15}$ $\Omega^{-1}$ $m^{-2}$) | $\lambda_{sf}$ (nm) for models 1 and 2, or $\Lambda$ for model 3 | $\delta$ |
|---|---|---|---|---|
| Ta/Py/Cu/Pd | 1 | 0.7± 0.14 | 4.8 ± 0.4 | 0 |
|  | 2 | 0.49 ± 0.12 | 2.79 ± 0.09 | (2±2)×$10^{-9}$ |
|  | 3 | 0.46 ± 0.56 | 0.2± 0.4 | 0.01± 0.68 |
| Ta/Py/Pd | 1 | 2.8± 0.5 | 5.2 ± 0.5 | 0.21 ± 0.02 |
|  | 2 | 1.33 ± 0.12 | 2.7 ± 0.4 | 0.24± 0.02 |
|  | 3 | 37 ± 278 | 0.3 ± 0.1 | 0.17 ± 0.15 |
| Ta/Py/Cu/Pt | 1 | 0.471 ± .009 | 1.18 ± 0.07 | 0.01± 0.02 |
|  | 2 | 0.44 ± .02 | 0.8 ± 0.4 | 0.36± 0.09 |
|  | 3 | 0.72± 0.05 | 0.18 ± .05 | (2±4)× $10^{-8}$ |
| Ta/Py/Pt | 1 | 4.9± 0.3 | 1.2 ± 0.1 | 0.30± 0.06 |
|  | 2 | 2.65± 0.11 | 0.37± 0.09 | 0.30± 0.07 |
|  | 3 | 12± 40 | 0.2 ± 0.3 | 0.42 ± 0.40 |

Table 2: Fitting parameters obtained by use of the three different models for spin relaxation rate vs. conductivity described in the text.

For all four sample types, Model 3 is a poor fit to the data, though it is marginally acceptable for Py/Cu/Pt with respect to the data error bars. Nevertheless, we see clearly that the assumption that both $\tau$ and $\tau_{sf}$ are unaffected by sample thickness does not agree with the qualitative dependence of $\alpha$ on $L_{NM}$, at least within the context of conventional diffusive spin-transport theory.

For samples with Cu spacer layers, Model 1 provides an adequate fit, but the accuracy of the fit to Model 2 is markedly improved. The interfacial spin-loss parameter is negligible for all cases with Pd. For Py/Cu/Pt(x), the fitted value for $\delta$ is no longer negligible.

Omission of the Cu spacer layer has a strong effect on the fitted parameters. The fits yield similar values of $\lambda_{sf}$ for Py/Pd(x) compared to Py/Cu/Pd(x), but $G^{\uparrow\downarrow}$ is substantially enhanced (3-4 times) for the Py/Pd interface relative to that of the Py/Cu interface. The fitting result for Model 3 is anomalous, with a nonphysical value for $G^{\uparrow\downarrow}$ when compared to first-principles calculations [74]. The fitted values for $\delta$ are no longer negligible for Py/Pd, suggesting that proximity polarization of Pd might be a source of interfacial spin-flip. For all three models, the fitted values for $\lambda_{sf}$ are substantially smaller than the bulk Pd mean-free-path of 23 nm.

As is the case for Py/Pd(x), the fits for Py/Pt(x) yielded similar values for $\lambda_{sf}$, a marked increase in the values for $G^{\uparrow\downarrow}$ compared to samples with a Cu spacer, and non-negligible results for $\delta$. Again, the latter result is consistent with the hypothesis that proximity polarization plays a role in the interfacial spin-flip process, though the significant value of $\delta$ for Py/Cu/Pt($x$) suggests that proximity polarization is not the only factor to consider. Again, the fitted value for $G^{\uparrow\downarrow}$ obtained with Model 3 is unphysically large, as seen in the case of Py/Pd($x$).

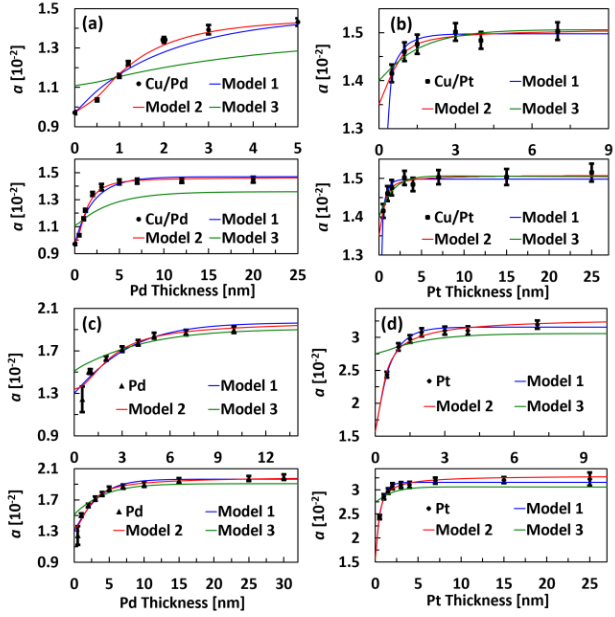

Figure 3: (a) Measured damping $\alpha$ (black dots) as a function of Pd thickness for Py/Cu/Pd($x$), together with fits to Models 1 (blue), 2 (red) and 3 (green). Top panel zooms in on smaller thicknesses while bottom panel shows data and fits over the entire range of Pd thicknesses that were measured. 3(b-d) is the same as 3(a), but for Py/Cu/Pt($x$), Py/Pd($x$), and Py/Pt($x$), respectively. In all four sample systems, Model 3, with an Elliot-Yafet-like proportionality between $\sigma$ and $\lambda_{sf}$, was grossly inadequate as a quantitative description of the data. Conversely, Model 2, with a Dyakanov-Perel-like proportionality between $\rho$ and $\lambda_{sf}$, provided the superior fitting results, though Model 1 is never inadequate.

This final result deserves some explanation. In the context of conventional diffusive spin-transport theory, the spin-pumping contribution to the damping in Model 3 for samples without a Cu spacer appears to be limited by the bulk spin-conductivity $\sigma/\lambda_{sf}$ of the NM, such that no amount of increase in the value of $G^{\uparrow\downarrow}$ can give rise to further increase in $\alpha$. Given that $\sigma \propto \lambda_{sf}$ in Model 3, the reduction in $\sigma$ for small NM thickness does not change the insensitivity of $\Delta\alpha$ to $G^{\uparrow\downarrow}$. Hence, the nonphysical fitted values of $G^{\uparrow\downarrow}$ for Model 3 are related to the general inadequacy of Model 3 to fit any of the data.

All the fits for NM = Pt yield $\lambda_{sf} < 1.5$ nm, a value that is vastly smaller than the bulk mean-free-path of $\ell_{Pt} = 16$ nm. Other workers have suggested that fits of the thickness-dependence for damping in spin-pumping measurements are subject to artifacts if interfacial spin-flip is not properly accounted for [75, 76]; by this argument, the spin-diffusion length for bulk Pt is actually longer than the fitted value, but the sensitivity of the fits to the bulk $\lambda_{sf}$ is negligible because the majority of spins-flips occur at the interface. This would lead to only a small net increase in the damping for NM thicknesses greater than that required to completely form the Py/NM interface. However, we still observe a small but statistically significant increase in $\alpha$ for Pt thicknesses up to 2 nm for both Py/Cu/Pt and Py/Pt, which is consistent with fitted values of $\lambda_{sf} < 1.5$ nm for Pt, even when interfacial spin-flip is accounted for. Hence, we conclude that the observed increase in damping with increasing Pt thickness indeed reflects a bulk spin-flip process, and not an interfacial effect. Our extracted value of λ$_{sf}$~0.8 nm for Pt is also consistent with recent fitting of first-principles calculations to the diffusive model [74].

To compare spin relaxation times $\tau_{sf}$ with momentum scattering times $\tau$ in the context of Model 2, we use values for the electronic density of states at the Fermi surface $\nu_{DOS}$, once again obtained via de Haas-van Alphen measurements for Pd and Pt: $\nu_{DOS,Pd} = 3.97$ eV$^{-1}$ atom$^{-1}$, $\nu_{DOS,Pt} = 2.79$ eV$^{-1}$ atom$^{-1}$ [71,72]. We use the Einstein relation for the diffusion parameter $D = 1/(\rho e^2 \nu_{DOS} n_{at})$, where $n_{at}$ is the atomic density, and the diffusion equation $\lambda_{sf} = \sqrt{D\tau_{sf}}$, as well as the fitted parameters obtained in Table 2 with Model 2, to calculate the dependence of the spin-relaxation time on NM thickness for all samples, as shown in Fig. 4.

For all samples with Pd, we find that the asymptotic fitted spin-relaxation time at large Pd thickness of $\cong 45$ fs is indeed longer than the momentum scattering time of $\cong 21$ fs, consistent with the supposition of an E-Y spin-flip process for bulk Pd. This is in contrast to the conclusion presented in Ref. [77], where Foros, *et al.* claimed the observation of a spin-diffusion length that was equal to the mean free path (~ 9 nm) was indicative of a paramagnon-mediated spin-decoherence process. However, the relative scale of the mean free path and the spin-diffusion length is not the appropriate comparison when attempting to ascertain the mechanics of spin-scattering: the relevant parameters are the spin scattering time and momentum scattering time [34, 42]. Indeed, as we show here, while the mean free path of $\cong 23$ nm is in actuality longer than the spin-diffusion length $\cong 2.8$ nm, the momentum-scattering time is still shorter than that for spin-scattering. The bulk spin lifetime in Pt, as calculated assuming the diffusive theory, is shorter than the bulk momentum scattering time, as seen in Fig. 4(b) and 4(d). The diffusive theory based on a Boltzmann treatment [78] requires that the spin lifetime be greater than the scattering time. For Pt, therefore, either the simple diffusive theory must be inadequate, requiring theoretical extensions, or the decay length observed in the damping and some inverse spin Hall measurements cannot be interpreted as a spin diffusion length. Further work is therefore needed to determine the dominant mechanism of spin transport in such a material.

The failure of the diffusive model for Pt does not discount the importance of this work in terms of measuring spin diffusion lengths and scattering times for more ordinary metals. Indeed, verification of spin diffusion lengths and interfacial spin loss parameters independent from spin valve measurements are needed, especially at room temperature [79]. Our method can also be extended to low temperatures to compare with lateral spin valve measurements similar to those in Ref. [46].

While the data of Model 2 clearly provide the best fit, the question remains as to how such phenomenology can be appropriate for isotropic, transition metal alloys, given that the Dyakanov-Perel mechanism is disallowed in these materials since they lack the necessary translational symmetry breaking. Possibly, the fact that the change in conductivity in our samples is dominated by interfacial and grain boundary scattering, which do break translational symmetry, could contribute to the Dyakanov-Perel-like nature of the observed dependences. Future work studying impurity scattering and thermal dependence can be used to check this hypothesis.

An alternative understanding of the data presented here is that the spin-relaxation rate might be reduced at smaller film thicknesses for reasons that are independent of the enhanced resistivity. For example, electron spin-relaxation rates can be reduced for extremely small (< 2-3 nm) metallic nanoparticles when measured with electron spin resonance [80,81,82]. Such a reduction is to be expected if the electron energy level spacing $\Delta$ is larger than the energy broadening due to spin-flip processes $\hbar/\tau_{sf}$ [83]. If the condition $\Delta > \hbar/\tau_{sf}$ is satisfied, the spin-flip rate is renormalized, such that

$$\frac{1}{\tau_{sf}} \approx \frac{1}{\tau_{sf}^{(0)}} \left(1 + \frac{\tau_{sf}\Delta}{\hbar}\right)^{-1}. \tag{11}$$

Given that the NM layer thickness is indeed thinner than the mean free path for most of the samples that we have measured, there is the possibility that lateral confinement has a nontrivial effect on both the momentum- and spin-relaxation processes. In the case of thin films, such considerations do not generally apply because the lateral component of the wave vector is not quantized. However, if the spin-flip process is highly anisotropic, such that the spin relaxation rate is strongly enhanced when scattering between momentum states perpendicular to the film plane, then quantization in the perpendicular direction could still lead to a suppression of the spin-relaxation rate. In other words, quantum confinement for a thin film can be important only if the intra-band scattering matrix element is zero, i.e., only momentum perpendicular to the film is lost in a spin-scattering event. In principle, this might be true if the spin-scattering process in the NM is renormalized by the spin-Hall effect, whereby a pure spin-current flowing perpendicular to the FM/NM interface is converted into a charge current flowing parallel to the interface. If this is true, in-plane and out-of-plane FMR measurements using our technique will result in different spin diffusion lengths for the same material. Both Pt and Pd exhibit a significant spin-Hall effect [63], which suggests that such speculation is not without physical basis. If we use the well-known result $\Delta = (\hbar^2 k_F \pi)/(m^* L)$ for one-dimensional confinement of a free electron at the Fermi level, where $L$ is the film thickness, then quantum confinement is important for $L < 4.3$ nm in the case of Pt, and $L < 74$ nm in case of Pd. The increase in $\tau_{sf}$ with decreasing Pt-thickness in Fig. 4 appears to occur close to the critical thickness for quantum confinement, but the estimate for Pd vastly overestimates what is observed.

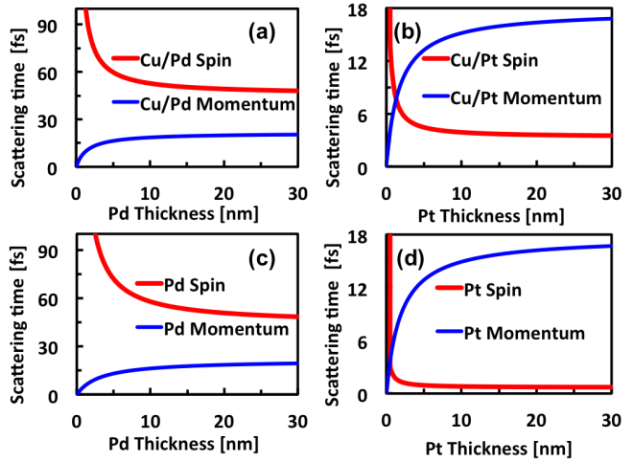

Figure 4: Calculated spin (red) and momentum (blue) scattering times for the experimental structures and extracted parameters from Model 2, for (a) Pd on Cu, (b) Pt on Cu, (c) Pd in direct contact with Py, and (d) Pt in direct contact with Py.

## D. Conclusions

We have measured spin- and charge-transport properties of Pd and Pt using FMR measurements. In particular, we have used the increase in damping as a result of spin-pumping to probe the flow of spin-current from the FM probe layer. We presented a matrix method to facilitate rapid formulation of the spin-back-flow parameter in arbitrary, one-dimensional multilayer stacks. We used precise measurements of the NM conductivity in conjunction with the FMR measurements to determine spin-diffusion length, the FM/NM spin-mixing conductance, and the interfacial spin-flip parameter, all in the context of three different models for the correlation of charge- and spin-transport properties. All the data are most accurately fitted with a model where $\tau_{sf} \propto 1/\tau$, in sharp contrast to the usual presumption that E-Y processes should dominate for spin-relaxation in transition metals. We confirm that the room-temperature Pt-spin-diffusion length is less than 1.5 nm for all samples measured. However, when we use the model with the D-P phenomenology to fit the data, we find that $\tau < \tau_{sf}$ for all the samples with Pd, consistent with the use of diffusive transport theory, but $\tau > \tau_{sf}$ for all samples with Pt except those with the very thinnest ($< 2$ nm) Pt layers. We confirmed that the imaginary part of the spin-mixing conductance for both the FM/Pt and FM/Pd interfaces is negligible, as evidenced by the absence of any systematic dependence of the gyromagnetic ratio on NM thickness. Given the success of fits based upon a D-P phenomenology, in spite of the fact that Dyakanov-Perel does not apply for materials with inversion symmetry, we suggest that theories for spin transport in such thin materials might benefit from the inclusion of quantum confinement and other low-dimensional considerations, insofar as the NM thickness is comparable to the electron mean-free-path in these materials, and the spin Hall effect might result in some degree of anisotropy for the spin-relaxation process that is not accounted for in our current theoretical understanding for diffusive spin transport.

We acknowledge helpful conversations with Mark Stiles and Mathias Weiler. CTB acknowledges support from the National Research Council Postdoctoral Research Associates program.

---

[1] I. Zutic, J. Fabian, and S. Das Sarma, *Rev. Mod. Phys.* 76, 323 (2004)
[2] J. Hirsch, *Phys. Rev. Lett.* 83, 1834 (1999)

[3] M.I. Dyakonov, Spin Hall Effect, arXiv:1210.3200;

[4] S. Emori, U. Bauer, S. M. Ahn, E. Martinez, G. S. D. Beach, *Nature Materials* 12, 611-616, (2013)

[5] Tomas Jungwirth, Jörg Wunderlich, Kamil Olejník, *Nature Materials* 11, 382 (2012)

[6] S. O. Valenzuela and M. Tinkham, *Nature* 442, 176–179 (2006)

[7] T. Kimura, et al, *Phys. Rev. Lett.* 98, 156601 (2007)

[8] A. Brataas, A. D. Kent, and H. Ohno, *Nature Materials* 11, 372-381 (2012).

[9] A. V. Khvalkovskiy, D. Apalkov, S. Watts, R. Chepulskii, R. S. Beach, A. Ong, X. Tang, A. Driskill-Smith, W. H. Butler, P. B. Visscher, D. Lottis, E. Chen, V. Nikitin, and M. Krounbi, *J. Phys. D: Appl. Phys.* 46, 074001 (2013).

[10] S. Yuasa, and D. D. Djayaprawira, *J. Phys. D: Appl. Phys.* 40, R337

[11] Yaroslav Tserkovnyak, Arne Brataas and Gerrit E. W. Bauer, *Phys. Rev. B* **66,** 224403 (2002)

[12] K. Lenz, et al, *Phys. Rev. B* 69, 144422 (2004)

[13] A. Brataas, Y. Terkovnyak, G. E. W. Bauer, and P. J. Kelly, *Spins Current*, Part 1, Ch. 8, ed. Sadamichi Maekawa, Sergio O. Valenzuela, Eiji Saitoh, and Takashi Kimura, Oxford University Press (2012)

[14] G. Woltersdorf, et al, *Phys. Rev. Lett.* 95, 037401 (2005)

[15] E. Saitoh, M. Ueda, H. Miyajima and G. Tatara, *Appl. Phys. Lett.* 88, 182509 (2006)

[16] M. V. Costache et al, *Phys. Rev. Lett.* 97, 216603 (2006)

[17] I. M. Miron, et al, *Nature,* 476, 189 (2011)

[18] L. Liu, Chi-Feng Pai, Y. Li, H. W. Tseng, D. C. Ralph, and R. A. Buhrman, *Science* 336, 555 (2012).

[19] L. Liu, R. A. Buhrman, and D. C. Ralph, arXiv:1111.3702 (2012).

[20] V. E. Demidov, et al, *Appl. Phys. Lett.* 99, 172501 (2011)

[21] V. E. Demidov, et al, *Nature Materials* 11, 1028–1031 (2012)

[22] P. P. J. Haazen, et al, *Nature Materials* 12, 299–303 (2013)

[23] Z. Duan, et al, *Phys. Rev. B* 90, 024427 (2014)

[24] Z. Duan, et al, *Nature Communications* 5, 5616 (2014)

[25] Y. Niimi, D. Wei, H. Idzuchi, T. Wakamura, T. Kato, And Y. Otani, *Phys. Rev. Lett.* 110, 016805 (2013)

[26] A. Azevedo, L.H. Vilela-Leao, R.L. Rodriguez-Suarez, A.F. Lacerda Santos, And S.M. Rezende, *Phys. Rev. B 83*, 144402 (2011)

[27] V. Castel, N. Vlietstra, J. Ben Youssef, And B.J. Van Wees, *Appl. Phys. Lett.* 101, 132414 (2012)

[28] W. Zhang, V. Vlaminck, J. E. Pearson, R. Divan, S. D. Bader, and A. Hoffmann, *Appl. Phys. Lett.*, 103, 242414 (2013)

[29] X. D. Tao, Z. Feng, B. F. Miao, L. Sun, B. You, D. Wu, J. Du, W. Zhang, H. F. Ding, *J. Appl. Phys.*, 115, 17C504 (2014)

[30] V. Vlaminck, J. E. Pearson, S. D. Bader, A. Hoffmann, *Phys. Rev. B* 88, 064414 (2013)

[31] S. Takahashi and S. Maekawa, *Phys. Rev. B* **67**, 052409 (2003)

[32] M. L. Polianski and P. W. Brouwer, *Phys. Rev. Lett.* **92**, 026602 (2004)

[33] Mark Johnson and R. H. Silsbee, *Phys. Rev. B* 37, 10 (1988)

[34] Y. Tserkovnyak, A. Brataas, G. E. W. Bauer, B. Halperin, *Rev. Mod. Phys.*, 77 (2005)

[35] M. I. Dyakonov and A.V. Khaetskii, "Spin Hall effect," in Spin Physics in Semiconductors, ser. in Solid-State Sciences, M. I. Dyakonov, Ed. New York: Springer, 2008, vol. 157, ch. 8, pp. 211–243

[36] C. T. Boone, H. T. Nembach, J. M Shaw, T.J. Silva, *J. Appl. Phys.*, 113, 153906 (2013)

[37] L. Vila, T. Kimura, Y. Otani, *Phys. Rev. Lett*. 99, 226604 (2007)

[38] R. J. Elliott, *Phys. Rev*. 96, 266 (1954)

[39] Y. Yafet, *Physics Letters A* 98, 287 (1983)

[40] M. Dyakonov and V. Perel, *Soviet Physics SolidState*, USSR 13, 3023 (1972)

[41] M. D. Mower, G. Vignale, I. V. Tokatly, *Phys. Rev. B* 83, 155205 (2011)

[42] H. Jiao and G.E.W. Bauer, *Phys. Rev Lett*. 110, 217602 (2013)

[43] D. Qu, S. Y. Huang, B. F. Miao, S. X. Huang, and C. L. Chien, *Phys. Rev. B* 89, 140407(R) (2014)

[44] J. Foros, G. Woltersdorf, B. Heinrich, and A. Brataas, *J. Appl. Phys.* **97**, 10A714 (2005).

[45] G. Mihajlovic, J. E. Pearson, S. D. Bader, and A. Hoffmann, *Phys. Rev. Lett.* 104, 237202

[46] M. Morota, Y. Niimi, K. Ohnishi, D. H. Wei, T. Tanaka, H. Kontani, T. Kimura, and Y. Otani, *Phys. Rev. B* 83, 174405 (2011)

[47] H. Kurt, R. Lolee, K. Eid, W. P. Pratt, J. Bass, Appl. Phys. Lett. 81, 4787 (2002)

[48] Kouta Kondou, Hiroaki Sukegawa, Seiji Mitani, Kazuhito Tsukagoshi, and Shinya Kasai, *Appl. Phys. Express* **5**, 073002 (2012).

[49] J. M. Shaw, H. T. Nembach, T.J. Silva, *Phys. Rev. B* 85, 054412 (2012)

[50] A. Ghosh, S. Auffret, U. Ebels, W. E. Bailey, *Phys. Rev. Lett*. 109, 127202 (2012)

[51] H. Y. T. Nguyen, R. Acharyya, E. Huey, B. Richard, R. Loloee, W. P. Pratt, Jr., J. Bass, Shuai Wang, and Ke Xia, *Phys. Rev. B* 82, 220401(R), (2010)

[52] O Gunnarsson, *J. Phys. F: Metal Phys*. Vol. 6, Num. 4 (1976)

[53] W. E. Bailey, A. Ghosh, S. Auffret, E. Gautier, U. Ebels, F. Wilhelm, and A. Rogalev, *Phys. Rev. B* 86, 14403 (2012)

[54] A. Ghosh, S. Auffret, U. Ebels, F. Wilhelm, A. Rogalev, W. E. Bailey, arXiv:1308.0450 (2013)

[55] O. Mosendz, J. E. Pearson, F. Y. Fradin, G. E. W. Bauer, S. D. Bader, and A. Hoffmann, *Phys. Rev. Lett*. 104, 046601 (2010)

[56] Wanjun Park, David V. Baxter, S. Steenwyk, I. Moraru, W. P. Pratt, Jr., and J. Bass, Phys. Rev. B 62, 1178 (2000)

[57] H. D. Liu, Y-P. Zhao, G. Ramanath, S. P. Murarka, G-C. Wang, *Thin Solid Films*, Vol 384, Issue 1, 151-156 (2001)

[58] A. T. McCallum and S. E. Russek, *Appl. Phys. Lett.* **84**, 3340 (2004).

[59] W. E. Bailey, Shan Wang, E. Y. Tsymbal, *Phys. Rev. B* 61, 1330 (2000)

[60] A. F. Mayadas and M. Shatzkes, *Phys. Rev. B* Vol. 1, Number 4, 1382 (1969)

[61] Justin M. Shaw, et al., Phys. Rev. B 80 184419 (2009)

[62] L. Liu, T. Moriyama, D. C. Ralph, and R. A. Buhrman, *Phys. Rev. Lett.* 106, 036601 (2011)

[63] O. Mosendz, V. Vlaminck, J. E Pearson, F. Y. Fradin, G. E. W. Bauer, S. D. Bader, and A. Hoffmann, *Phys. Rev. B* 82, 214403 (2010)

[64] Th. Gerrits, M. L. Schneider, T. J. Silva, *J. Appl. Phys*. 99, 023901 (2006)

[65] H. T. Nembach, T. J. Silva, J. M. Shaw, M. L. Schneider, M. J. Carey, and J. R. Childress, *Phys. Rev. B* 84, 054424 (2011)

[66] R. D. McMichael and P. Krivosik, *IEEE Trans. Magn.* 40, 2 (2004)

[67] J. M. Shaw, H. T. Nembach, T. J. Silva, C. T. Boone, *J. Appl. Phys.* 114, 243906 (2014)

[68] A. Starikov, P.J. Kelly, A. Brataas, Y. Tserkovnyak, and G.E.W. Bauer, *Phys. Rev. Lett.* 105, 236601 (2010)

[69] L.L. Henry, Q. Yang, W.-C. Chiang, P. Holody, R. Loloee, W. P. Pratt, Jr., and J. Bass, *Phys. Rev. B* 54, 12336 (1996)

[70] K. Xia, P. J. Kelly, G. E. W. Bauer, I. Turek, J. Kudrnovsky, and V. Drchal, *Phys. Rev. B* 63, 064407 (2001)

[71] L. R. Windmiller, J. B. Ketterson and S. Hornfeldt. *Phys. Rev. B*, Vol. 3, Number 12, 4213 (1971)

[72] J. B. Ketterson and L. R. Windmiller. *Phys. Rev. B*, Vol. 2, Number 12, 4812 (1970)

[73] M. Marder, *Condensed Matter Physics*, John Wiley & Sons (2000)

[74] Y. Liu, et al., Phys. Rev. Lett. 113, 207202 (2014)

[75] J.-C. Rojas-Sánchez, N. Reyren, P. Laczkowski, W. Savero, J.-P. Attané, C. Deranlot, M. Jamet, J.-M. George, L. Vila, and H. Jaffrès, *Phys. Rev. Lett.* 112, 106602 (2014)

[76] H. Y Nguyen, J. Pass, W. Pratt, arXiv:1310.4364

[77] J. Foros, G. Woltersdorf, B. Heinrich, and A. Brataas, *J. Appl. Phys.* 97, 10A714 (2005)

[78] T. Valet and A. Fert, Phys. Rev. B 48, 7099 (1993)

[79] Jack Bass and William P. Pratt, *J. Phys.: Condens. Matter* 19 183201 (2007)

[80] *Physical Status Solidi b*, vol. 24, 525 (1967)

[81] *Journal De Physique*, vol. C2, 115 (1977)

[82] *J. Phys. Colloques*, vol. 38, 109 (1977)

[83] A. Kawabata, *J. Phys. Soc. Jap.* Vol. 29, No. 4, (1970)